\documentclass{emulateapj}

\begin{document}

\shortauthors{Luhman et al.}
\shorttitle{Wide Binary Brown Dwarf}

\title{Discovery of a Wide Binary Brown Dwarf Born in Isolation\altaffilmark{1}}

\author{
K. L. Luhman\altaffilmark{2,3},
E. E. Mamajek\altaffilmark{4},
P. R. Allen\altaffilmark{2},
A. A. Muench\altaffilmark{5},
and D. P. Finkbeiner\altaffilmark{5}
}

\altaffiltext{1}
{Based on observations performed with the Magellan Telescopes 
at Las Campanas Observatory and the {\it Spitzer Space Telescope}.}

\altaffiltext{2}{Department of Astronomy and Astrophysics, The Pennsylvania
State University, University Park, PA 16802; kluhman@astro.psu.edu.}

\altaffiltext{3}{Visiting Astronomer at the Infrared Telescope Facility, which
is operated by the University of Hawaii under Cooperative Agreement no. NCC
5-538 with the National Aeronautics and Space Administration (NASA), 
Office of Space Science, Planetary Astronomy Program.}

\altaffiltext{4}{Department of Physics and Astronomy,
The University of Rochester, Rochester, NY 14627.}

\altaffiltext{5}{Harvard-Smithsonian Center for Astrophysics, Cambridge, 
MA 02138.}

\begin{abstract}

During a survey for stars with disks in the Taurus star-forming region using 
the {\it Spitzer Space Telescope}, we have discovered a pair of young brown 
dwarfs, FU~Tau~A and B, in the Barnard 215 dark cloud. 
They have a projected angular separation of $5.7\arcsec$, 
corresponding to 800~AU at the distance of Taurus. 
To assess the nature of these two objects, we have obtained spectra of 
them and have constructed their spectral energy distributions. 
Both sources are young ($\sim1$~Myr) according to their H$\alpha$ emission,
gravity-sensitive spectral features, and mid-IR excess emission.
The proper motion of FU~Tau~A provides additional evidence of its membership
in Taurus.  We measure spectral types of M7.25 and M9.25
for FU~Tau~A and B, respectively, which correspond to masses of $\sim$0.05 
and $\sim$0.015~$M_\odot$ according to the evolutionary models of 
Chabrier and Baraffe.
FU~Tau~A is significantly overluminous relative to an isochrone passing 
through FU~Tau~B and relative to other members of Taurus near its spectral 
type, which may indicate that it is an unresolved binary.  
FU~Tau~A and B are likely to be components of a binary system based on 
the low probability ($\sim3\times10^{-4}$) that Taurus would produce two 
unrelated brown dwarfs with a projected separation of $a\leq6\arcsec$.
Barnard 215 contains only one other young star and is in a remote 
area of Taurus, making FU~Tau~A and B the first spectroscopically-confirmed
brown dwarfs discovered forming in isolation rather than in a stellar cluster
or aggregate. Because they were born in isolation and comprise a weakly bound 
binary, dynamical interactions with stars could not have played a role 
in their formation, and thus are not essential for the birth of brown dwarfs. 

{\bf ERRATUM:} The $K$-band magnitude for FU~Tau~B in Table~1 is incorrect and
should be
13.33. The bolometric luminosity of FU~Tau~B in Table 3 and Figure 5 is
incorrect because of that mistake and a separate arithmetic error.
The correct value of the luminosity is 0.0039~$L_\odot$.
FU~Tau~A and B exhibited different isochronal ages in the
original Hertzsprung-Russell diagram in Figure~5, which was unexpected
for members of a binary system. This discrepancy is reduced in the
corrected version of Figure~5 since both objects are now above the
isochrone for 1~Myr.
Given the large uncertainties in model isochrones at such young ages, the 
positions of FU~Tau~A and B in Figure~5 could be roughly consistent
with coevality.

\end{abstract}


\section{Introduction}
\label{sec:intro}

Several theoretical studies have suggested that dynamical interactions 
among young stars could be important -- perhaps even essential -- for the 
formation of brown dwarfs \citep{rc01,bos01,bat02,del03,umb05,goo07,sta07}.
In one of the proposed scenarios, the dynamical evolution of a group
of protostars leads to the ejection of one of its members from the natal cloud
core. Because its accretion is prematurely halted, the ejected object does
not grow to become a star. Alternatively, brown dwarfs might form through
the fragmentation of circumstellar disks around protostars. 
The disruption of these binary systems by interactions with other stars
in the surrounding cluster would then produce free-floating brown dwarfs. 

According to early versions of the ejection models, binary brown dwarfs 
should have separations that are no larger than $\sim10$~AU \citep{bat02}.
This prediction was consistent with the results from initial multiplicity 
surveys of brown dwarfs \citep[][references therein]{bur07}, but more recent 
observations have uncovered a small number of wide low-mass binaries in 
both the field and young clusters. Because of their fragile nature, it would 
seem difficult for a dynamical model to explain the existence of the widest 
and least massive systems \citep[$a\gtrsim100$~AU, 
$M_{tot}\lesssim0.15$~$M_\odot$;][]{whi99,cha04,luh04bin,bil05,all06,jay06,
cab06,cab07,clo07,art07,bej08}.
Nevertheless, in a new set of simulations, \citet{bat05} were able to 
produce a wide binary brown dwarf through the simultaneous ejection of 
two brown dwarfs in similar directions, indicating that the ejection models 
may remain viable.
However, the feasibility of this formation mechanism for wide binaries
depends on the environmental conditions of the star-forming cloud, particularly
the stellar density. 
In fact, because the ejection models require the presence of a stellar cluster, 
brown dwarfs should not form in isolation, either as singles or binaries, if 
dynamical interactions are necessary for their formation.

The Taurus complex of molecular clouds is the most promising site 
in which one might find brown dwarfs that have been born in isolation.
Taurus is the prototypical example of a low-density star forming 
region, is well-populated ($\sim400$ known members) and nearby ($d=140$~pc),
and has been surveyed extensively for both stellar 
and substellar members \citep{ken08}. 
In this paper, we report the discovery of a wide binary brown dwarf
in an isolated dark cloud in Taurus. 
We present optical and infrared (IR) photometry and spectroscopy for the 
components of the pair (\S~\ref{sec:obs}) and use these data to
measure their spectral types (\S~\ref{sec:class}), place them on the 
Hertzsprung-Russell (H-R) diagram (\S~\ref{sec:hr}), and construct their 
spectral energy distributions (\S~\ref{sec:sed}).
We then assess the evidence that these two objects are members of Taurus
(\S~\ref{sec:mem}) and that they comprise a binary system (\S~\ref{sec:bin}).
Finally, we discuss the implications of this new binary system for the
origin of brown dwarfs.
(\S~\ref{sec:disc}).

\section{Observations}
\label{sec:obs}

\subsection{Background on FU Tau}
\label{sec:background}

\begin{figure}
\epsscale{1}
\plotone{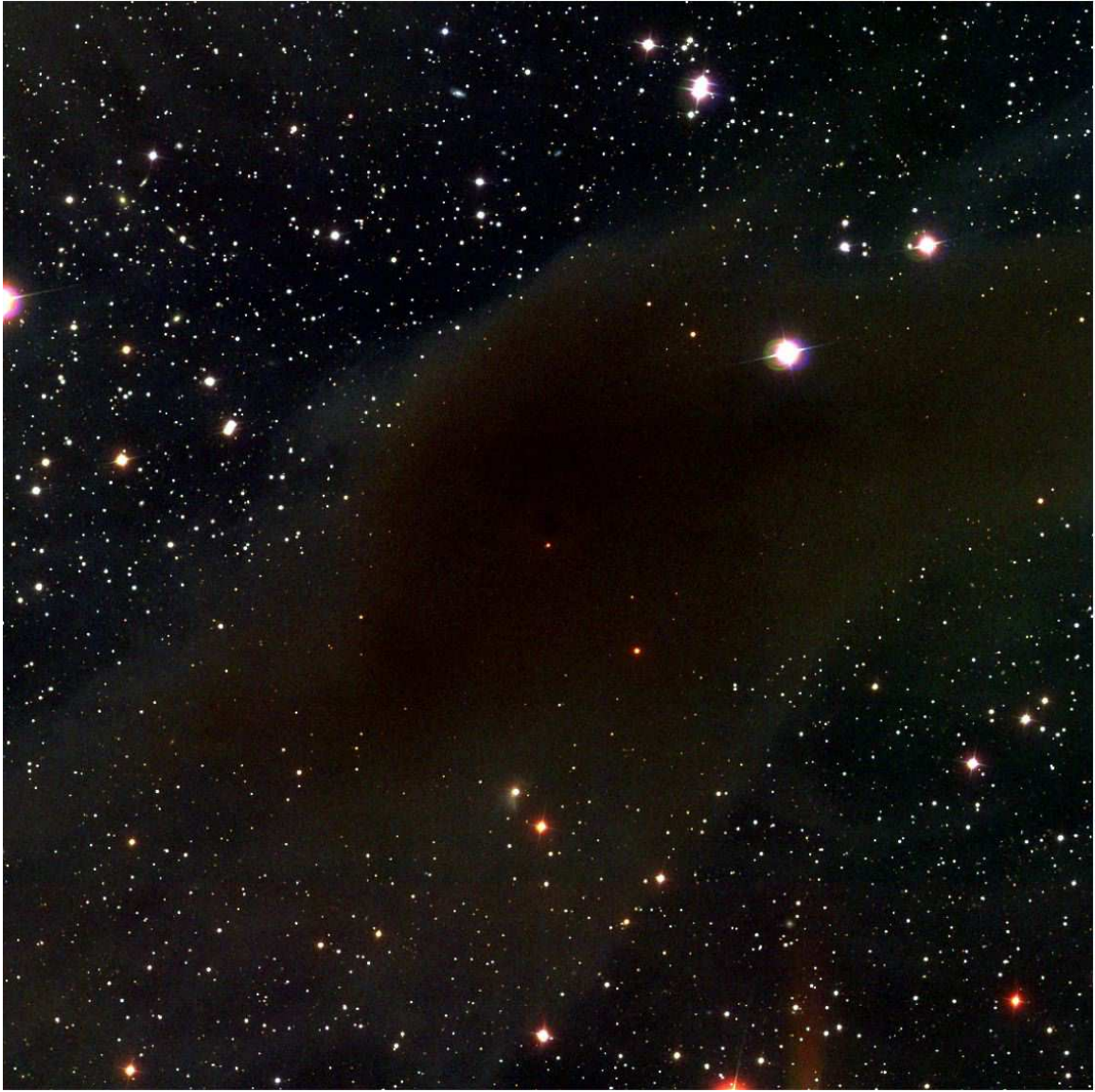}
\caption{
Optical image of the Barnard 215 dark cloud obtained by SDSS \citep{fin04}.  
FU~Tau~A and B are at the center of the image. The only other known member
of Taurus within this field is FT~Tau, which is the star surrounded 
by extended emission in the lower middle of the image. 
The red, green, and blue image planes correspond to the $i$, $r$, and $g$
filters, respectively. The size of the image is $0.5\arcdeg\times0.5\arcdeg$. 
North is up and East is left
}
\label{fig:sdss}
\end{figure}

\begin{figure}
\epsscale{0.55}
\plotone{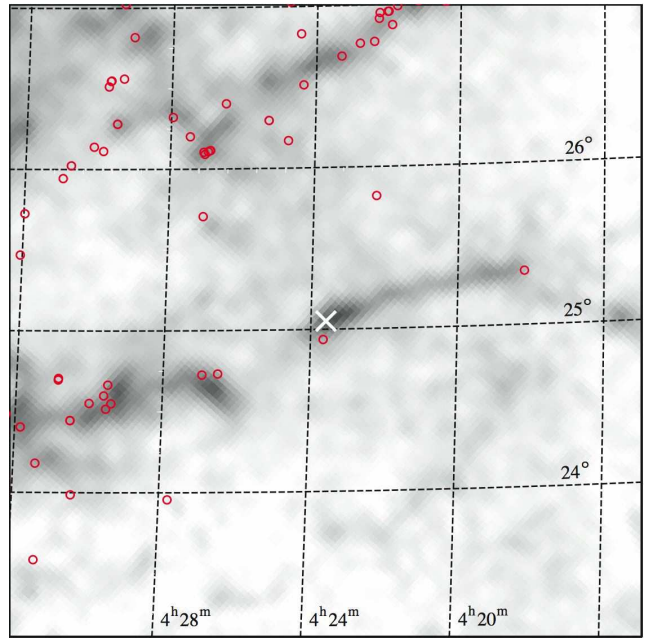}
\caption{
Extinction map from \citet{dob05} for a $2\arcdeg\times2\arcdeg$ field in the 
Taurus star-forming region centered on FU~Tau~A and B ({\it cross}). 
The positions of known members of Taurus are indicated ({\it circles}).
}
\label{fig:map}
\end{figure}

The subject of this study is the star FU~Tau, which is projected 
against the center of the Barnard 215 dark cloud 
\citep{bar27}\footnote{Additional designations for this cloud include
L1506E and L1506A \citep{lyn62,ner88,lm99}.}.
An optical image of FU~Tau and Barnard 215 is shown in Figure~\ref{fig:sdss}.
The cloud lies in a remote area of Taurus that is well-removed from 
most of the known members of the star-forming region, as illustrated
by the map in Figure~\ref{fig:map}. 
FU~Tau was first identified as a possible young star by \citet{har53} 
through an H$\alpha$ objective prism survey (designated as Haro~6-7). 
As indicated by its name, FU~Tau has exhibited significant variability 
\citep{kho98}, which is a common characteristic of young stars. 
\citet{jh79} identified it as a probable member of Taurus based on its proper 
motion. Despite the early evidence of the youth and membership of FU~Tau, 
no other significant work has been done on it. 

\citet{luh06tau2} used mid-IR images from the {\it Spitzer Space Telescope}
\citep{wer04} to search for new members of Taurus that have circumstellar 
disks.  During a continuation of that survey, we identified FU~Tau as a 
candidate based on its mid-IR excess emission. While performing optical 
spectroscopy on it, we noticed a fainter nearby object in the 
spectrometer's acquisition images. Because FU~Tau is projected 
against the center of a dark cloud, an optically visible background star in 
close proximity to it was unexpected. 
Therefore, we elected to obtain a spectrum of this object to determine if it is
a member of the cloud and a potential companion to FU~Tau. 
The spectra confirmed that both FU~Tau and the nearby source are young
and have late spectral types.
Hereafter in this paper, we refer to these objects as FU~Tau~A and FU~Tau~B.
We describe our spectroscopic observations of the pair in \S~\ref{sec:spec}.
To obtain photometry and astrometry for FU~Tau~A and B, 
we have searched the public data archives of various observatories and 
wide-field surveys for optical and IR images that encompass their location.
The data produced by this search are presented in \S~\ref{sec:image}.

\subsection{Spectroscopy}
\label{sec:spec}

We obtained long-slit optical spectra of FU~Tau~A and B on the nights of 
2007 December 17 and 18, respectively, using the Low Dispersion Survey 
Spectrograph (LDSS-3) on the Magellan~II Telescope. The observations were
performed with a 1.1$\arcsec$ slit, which was rotated to the parallactic angle. 
We used the VPH All and VPH Red grisms on the first and second nights, 
respectively, resulting in spectral resolutions of 10 and 5.5~\AA\ at 7500~\AA.
We collected one 10~min exposure for the primary and three 20~min exposures 
for the secondary. After bias subtraction and flat-fielding, we 
extracted the spectra and wavelength calibrated them with arc lamp data. 
We then corrected the spectra for the sensitivity functions of the detectors, 
which were measured from observations of a spectrophotometric standard star.

FU~Tau~A was also observed at near-IR wavelengths using SpeX \citep{ray03} 
at the NASA Infrared Telescope Facility.
On 2007 December 3, we operated SpeX in the prism mode with a $0.8\arcsec$ 
slit, which produced a spectrum that extended from 0.8--2.5~\micron\ and
that exhibited a resolving power of $R=100$. On 2007 December 26, we collected
another spectrum across the same wavelength range at higher resolution 
($R=1000$) by using SpeX in the SXD mode with a $0.8\arcsec$ slit.
These data were reduced with the Spextool package \citep{cus04} and corrected 
for telluric absorption \citep{vac03}. 

\subsection{Images}
\label{sec:image}

\subsubsection{2MASS}

Both components of FU~Tau are detected in images at $J$, $H$, and $K_s$ from 
the Two-Micron All-Sky Survey \citep[2MASS;][]{skr06}. 
FU~Tau~A and B correspond to 2MASS~J04233539+2503026 and J04233573+2502596,
respectively, in the 2MASS Point Source Catalog. 
The 2MASS photometry of the primary should have negligible contamination from
its companion given the large flux ratio of the pair at near-IR wavelengths 
($F_A/F_B\sim40$). Those data are provided in Table~\ref{tab:phot}.
Because the secondary is only marginally resolved by 2MASS, the
photometry reported in the Point Source Catalog may not be reliable. 
We do not use the 2MASS images for measuring the relative positions of FU~Tau~A 
and B since they are better resolved by data from other facilities. 

\subsubsection{SDSS}

Although it focused on areas of the sky at high galactic latitude, the 
Sloan Digital Sky Survey \cite[SDSS;][]{yor00} did obtain images of 
additional fields closer to the galactic plane \citep{fin04}. Those data 
encompassed a large portion of the Taurus cloud complex, including Barnard 215
(see Fig.~\ref{fig:sdss}). 
We have retrieved images and photometry of FU~Tau in the five optical bands 
of SDSS \citep[$ugriz$;][]{fuk96} from the Sixth Data Release of the survey 
\citep{ade08}. The calibration of these images is described by \citet{pad08}.
Two sets of images are available for FU~Tau, which were taken on 2002 
December 6 and 31. The FWHM is $\sim1\arcsec$ for point sources in 
all bands at both epochs.
The components of FU~Tau are well-resolved from each other in these data.
The primary was detected in each of the five bands while the secondary 
appeared in $r$, $i$, and $z$. 
A variety of flux measurements are provided by the Sixth Data Release. 
For FU~Tau~A and B, we have selected the data measured with an aperture
radius of $1.745\arcsec$, which is small enough to avoid contamination 
from the primary in the measurement of the secondary. Using other stars
in these images, we measured aperture corrections for each band between 
radii of $1.745\arcsec$ and $7.43\arcsec$ and applied these corrections
to the $1.745\arcsec$ data for FU~Tau~A and B. 
For the photometric errors, we combine the Poisson errors in the source and 
background emission with uncertainties of $\sim5$\% in the absolute 
calibrations.
The SDSS photometry and relative astrometry for the components of FU~Tau
are listed in Tables~\ref{tab:phot} and \ref{tab:astro}, respectively.

\subsubsection{CFHT}

\begin{figure}
\epsscale{0.55}
\plotone{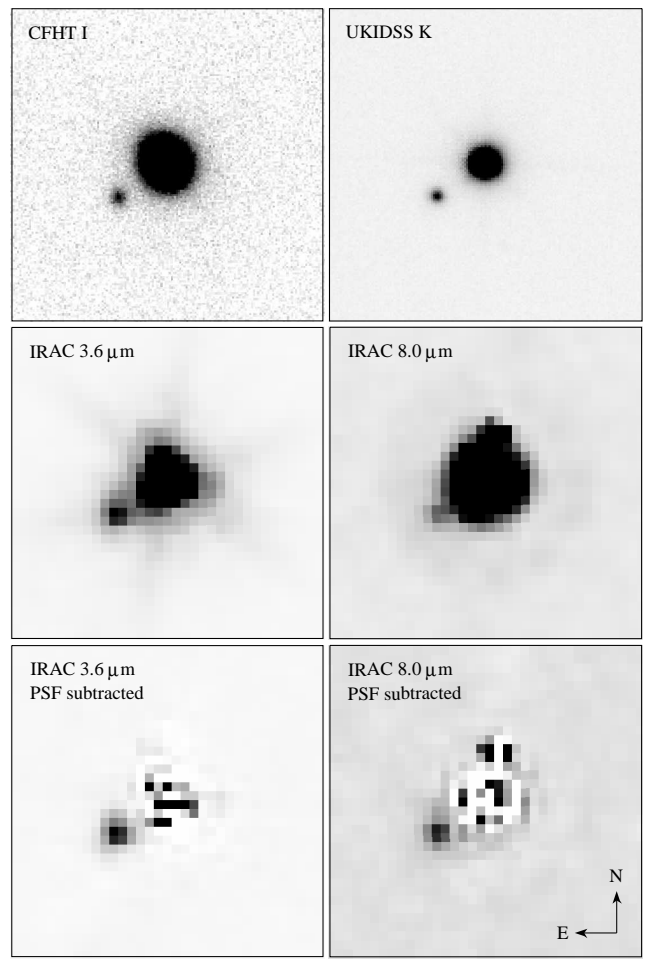}
\caption{
Optical and IR images of FU~Tau~A and B. The size of each image is 
$30\arcsec\times30\arcsec$.
}
\label{fig:image}
\end{figure}

\citet{gui06} obtained images at $I$ and $Z$ of the Barnard 215 cloud
as a part of a wide-field survey of Taurus. These data were collected
on the Canada-France-Hawaii Telescope (CFHT) with the CFH12K camera on
2002 December 29. The instrument contained twelve $2048\times4096$ CCDs 
in a $6\times2$ mosaic. The total field of view was $42\arcmin\times28\arcmin$.
We retrieved pipeline processed images of Barnard 215 and the associated 
photometric standards from the CFHT archive. 
One 30~s exposure and six 300~s exposures were available in each filter.
The FWHM of stars in the $I$-band images is 1--$1.2\arcsec$. 
One of the long $I$-band exposures of FU~Tau is shown in 
Figure~\ref{fig:image}. We measured aperture photometry for the primary 
and secondary from the short and long $I$-band exposures, respectively. 
The photometry at $I$ and relative astrometry are given in 
Tables~\ref{tab:phot} and \ref{tab:astro}, respectively.
Because the photometric standards for these observations lack photometry at 
$Z$ \citep{lan92}, we have not attempted to measure photometry from the 
images in this filter.

\subsubsection{UKIDSS}

The United Kingdom Infrared Telescope (UKIRT) Infrared Deep Sky Survey 
\citep[UKIDSS,][]{law07}\footnote{
UKIDSS uses the UKIRT Wide Field Camera \citep[WFCAM,][]{cas07} and a
photometric system described by \citet{hew06}. The pipeline processing and
science archive are described by Irwin et al. (in preparation) and
\citet{ham08}.} is obtaining near-IR images of large areas
of the northern sky. As a part of a survey of young nearby clusters,
UKIDSS is in the process of imaging most of the Taurus star-forming region in 
$Z$, $Y$, $J$, $H$, and $K$. The $K$-band observations of Barnard 215
were completed on 2005 December 31 and the resulting data products are now
available to the public. We retrieved the $K$-band image and photometry
for FU~Tau from the first UKIDSS data release \citep{war07}.
Point sources in this image exhibit a FWHM of $\sim0.9\arcsec$. The components 
of the binary are well-resolved, as shown in Figure~\ref{fig:image}. 
For a given object and band, UKIDSS provides several photometric measurements 
that are based on a range of aperture radii. We present in Table~\ref{tab:phot}
the $K$-band photometry for FU~Tau~B that was measured with an aperture radius 
of $1\arcsec$ \citep{dye06}. UKIDSS photometry for the primary is unavailable 
since it is saturated.  We measured the relative positions of the components of 
FU~Tau from the UKIDSS image, which are given in Table~\ref{tab:astro}.

\subsubsection{Spitzer Space Telescope}

As described in \S~\ref{sec:background}, our study of FU~Tau 
began during a survey for young stars with disks using mid-IR images of 
Taurus from the {\it Spitzer Space Telescope}. 
These images were obtained at 3.6, 4.5, 5.8, and 8.0~\micron\ with 
{\it Spitzer}'s Infrared Array Camera \citep[IRAC;][]{faz04}. 
FU~Tau appears within two sets of IRAC data.
On 2005 February 23, the young star FT~Tau, which is on the edge of 
Barnard 215 (see Fig.~\ref{fig:sdss}), was observed through the 
IRAC Guaranteed Time Observations of G. Fazio in {\it Spitzer} program 37. 
While the 3.6 and 5.8~\micron\ detectors were centered on FT~Tau, the 4.5 and 
8.0~\micron\ detectors observed an adjacent area of sky that happened
to encompass FU~Tau. The Astronomical Observation Request (AOR) identification
for these data is 3964672. FU~Tau also fell within a wide-field IRAC mosaic
that was obtained through the {\it Spitzer} Legacy program of
D. Padgett, which has a program identification of 30816. These observations 
were performed on 2007 March 30 and have AOR numbers of 19028480 and 19028224.

The IRAC images from 2005 and 2007 were processed with the {\it Spitzer} 
Science Center (SSC) S14.0.0 and S15.3.0 pipelines, respectively.
The images produced by the pipelines were then combined using R. Gutermuth's 
WCSmosaic IDL package. The reduced images of FU~Tau~A and B at 3.6 and 
8.0~\micron\ from the 2007 observations are shown in Figure~\ref{fig:image}. 
We measured aperture aperture photometry for FU~Tau~A
in the manner described by \citet{luh08cha}. We applied the same methods
to the secondary except that we first removed light from the 
primary by subtracting a scaled IRAC point spread function \citep{mar06}. 

In addition to the IRAC data, images of FU~Tau at 24~\micron\ have been
obtained with the Multiband Imaging Photometer for {\it Spitzer} 
\citep[MIPS;][]{rie04}. These observations were performed on 
2007 February 2 through D. Padgett's {\it Spitzer} Legacy program and 
have AOR numbers of 19026688 and 19027200.
We measured aperture photometry for FU~Tau~A from the images produced
by the SSC S16.1.0 pipeline \citep{luh08cha}. 
FU~Tau~B is not resolved from the primary in these data. 

The IRAC and MIPS measurements for FU~Tau are presented in Table~\ref{tab:phot}.

\section{Analysis}

\subsection{Spectral Classification}
\label{sec:class}

\begin{figure}
\epsscale{1}
\plotone{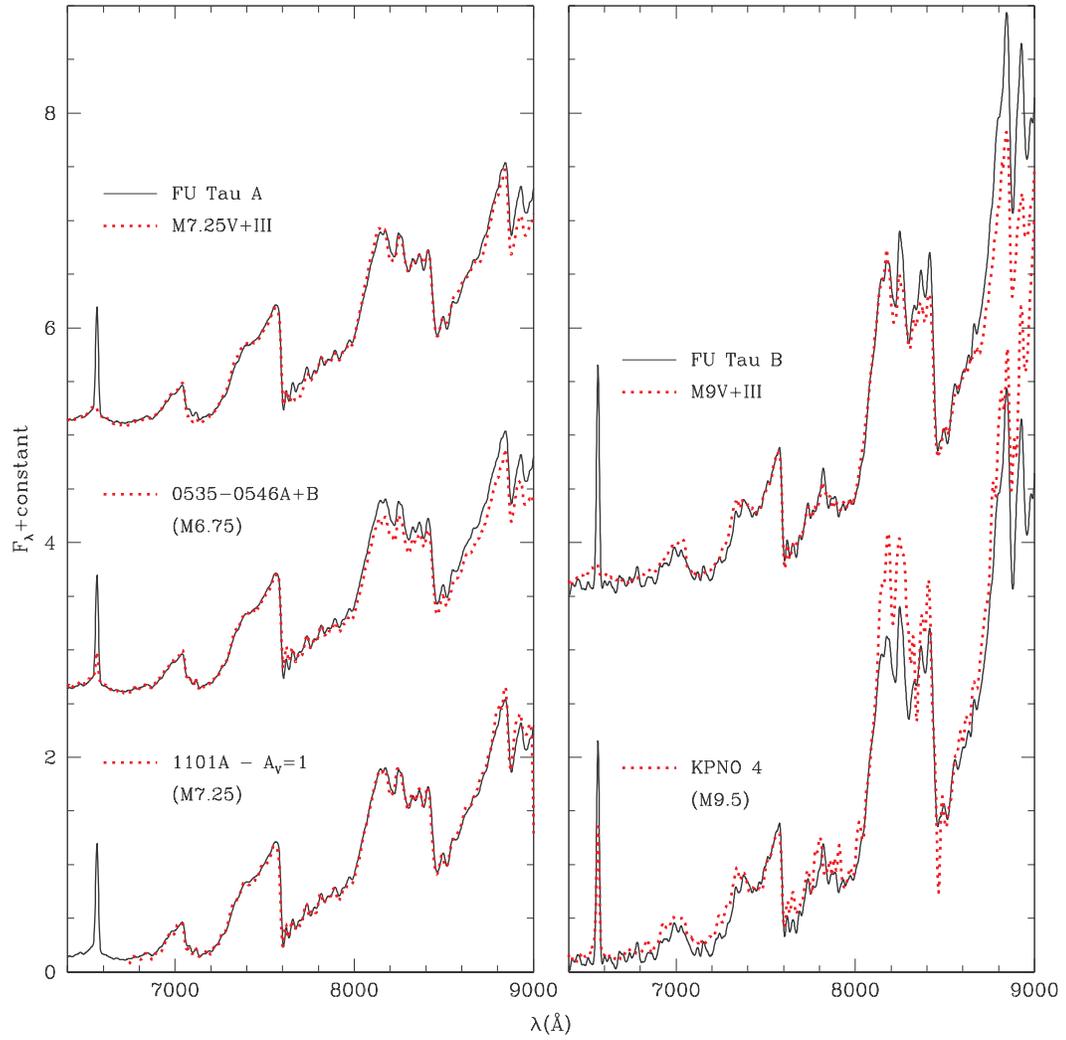}
\caption{
Optical spectra of FU~Tau~A and B ({\it solid lines}) compared to data for 
pre-main-sequence and field standards ({\it dotted lines}). {\it Left}:
After comparing FU~Tau~A to averages of standard dwarfs and giants,
we find that M7.25 provides the best match.
Its spectrum is slightly later than the composite spectrum of the 
eclipsing binary 2M~0535$-$0546 \citep[0.034 and 0.054~$M_\odot$;][]{sta06}
and is very similar to the primary in the young binary 2M~1101$-$7732 
\citep[M7.25;][]{luh04bin}.
{\it Right}: FU~Tau~B is slightly later than an average of M9 dwarfs
and giants and is earlier than the Taurus member KPNO~4
\citep[M9.5;][]{bri02}, leading to a classification of M9.25. 
The data are displayed at a resolution of 18~\AA\ and are normalized at
7500~\AA.
}
\label{fig:spec}
\end{figure}

To investigate the properties of FU~Tau~A and B, we begin by using our optical 
and IR spectra to measure their spectral types and reddenings and to assess 
their ages. The optical spectra are shown in Figure~\ref{fig:spec}. 
Both objects exhibit strong TiO and VO absorption bands, which are 
characteristic of late-M spectral types.
The weak Na~I, K~I, and FeH absorption lines \citep{mar96,luh98,gor03,mc04} 
and, in the case of the primary, triangular $H$-band continuum \citep{luc01}
demonstrate that FU~Tau~A and B have the low surface gravities that 
are found in pre-main-sequence objects.  
The H$\alpha$ emission in each spectrum is also stronger than that found 
among field dwarfs \citep{giz02}, providing additional evidence of youth. 
Thus, the spectra confirm that FU~Tau~A and B are young, low-mass objects. 

We measured spectral types from the optical spectra by comparing them to
averages of dwarfs and giants \citep{luh99} and previously-classified members 
of Taurus and other star-forming regions \citep{bri02}, arriving at 
M7.25$\pm$0.25 and M9.25$\pm$0.25 for the primary and secondary, respectively. 
To illustrate the derivation of these classifications, FU~Tau~A and B 
are compared to a selection of standards in Figure~\ref{fig:spec}.
FU~Tau~A matches closely with the average of a dwarf and a giant at M7.25 as 
well as the primary in the young binary 2MASS~J11011926$-$7732383 
\citep[hereafter 2M~1101$-$7732, M7.25;][]{luh04bin}. 
It is slightly cooler than the composite of the young eclipsing binary 
2MASS~J05352184$-$0546085 \citep[hereafter 2M~0535$-$0546;][]{sta06}, 
which has an optical spectral type of M6.75 \citep{luh07oph}.
Meanwhile, FU~Tau~B is intermediate between an M9 dwarf/giant
average and the Taurus member KPNO~4 \citep[M9.5;][]{bri02}.
When we compare the IR spectrum of FU~Tau~A to previous SpeX 
data for optically-classified young objects \citep{mue07}, we 
derive an IR spectral type that is consistent with the optical measurement.

The slopes of the optical spectra of FU~Tau~A and B are similar to those
of our bluest young standards, indicating they have low extinction ($A_V<1$).
For instance, the optical spectrum of the primary has the same slope
as Oph~1622$-$2405~A from Upper Sco \citep[M7.25;][]{luh07oph}, which 
probably has little reddening since it is not associated with a 
molecular cloud. However, the 1--2.5~\micron\ spectrum of FU~Tau~A is redder 
than that of Oph~1622$-$2405~A by an amount that is equivalent to 
$A_V=2$ \citep{rl85}. The same is true when we compare
FU~Tau~A to other young M7 objects that appear to have negligible extinction,
such as MHO~4, KPNO~2, and KPNO~5. This anomalous difference between the 
optical and IR reddening estimates for FU~Tau~A is probably not caused by 
errors in the 
spectra because the 0.6--1~\micron\ slopes of the LDSS-3, low-resolution SpeX
data, and medium-resolution SpeX data are in good agreement while the 
1--2.5~\micron\ slopes are the same among the two sets of SpeX data and 
the colors from 2MASS. It is possible that the IR spectrum appears too red
because it is contaminated by long-wavelength emission from circumstellar dust, 
or the optical spectrum appears too blue because it is contaminated by 
scattered light or UV emission from accretion. To test the first 
hypothesis, we have compared the $R=1000$ $K$-band spectrum of FU~Tau~A 
to SpeX data for field dwarfs near the same spectral type \citep{cus04}. 
We find no evidence for significant continuum emission from dust, which
would cause the lines to appear weaker, or ``veiled", relative to those
of a normal stellar photosphere. Thus, the slope of the IR spectrum 
of FU~Tau~A should accurately reflect its extinction. 
The alternative explanation is supported by the analysis of the spectral energy
distribution (SED) of FU~Tau~A in \S~\ref{sec:sed}, which demonstrates
the presence of short-wavelength excess emission.

\subsection{H-R Diagram}
\label{sec:hr}

To examine the masses and ages of the components of FU~Tau, we 
can compare their effective temperatures and bolometric luminosties to the 
values predicted by theoretical evolutionary models. 
We have converted the spectral types of FU~Tau~A and B to effective 
temperatures with the temperature scale from \citet{luh03ic}.
To estimate the luminosity of the primary, we have combined its $J$-band 
magnitude with the average bolometric correction for dwarfs near its 
spectral type \citep{dah02}, an extinction of $A_V=2$ (\S~\ref{sec:class}), 
and a distance of 140~pc \citep{wic98}. For the secondary, we use its $K$-band 
magnitude since a reliable $J$ measurement is not available. By doing so, 
we are assuming that any circumstellar disk that might reside around FU~Tau~B
produces negligible emission at $K$ compared to the stellar photosphere. 
This assumption is likely to be valid since disks around brown dwarfs rarely 
exhibit significant $K$-band excess emission \citep{luh08cha}.
The resulting temperatures and luminosities are listed in Table~\ref{tab:prop}. 
The uncertainties in $A_V$, near-IR magnitudes, and bolometric corrections
($\sigma\sim0.14$, 0.02, 0.1) correspond to errors of $\pm0.07$ in the 
relative values of log~$L_{\rm bol}$. When an uncertainty in the distance 
modulus is included ($\sigma\sim0.2$), the total uncertainties are $\pm0.11$.
These uncertainties do not include variability, which is known to be
significant for the primary \cite[Table~\ref{tab:phot};][]{kho98}.

\begin{figure}
\epsscale{0.55}
\plotone{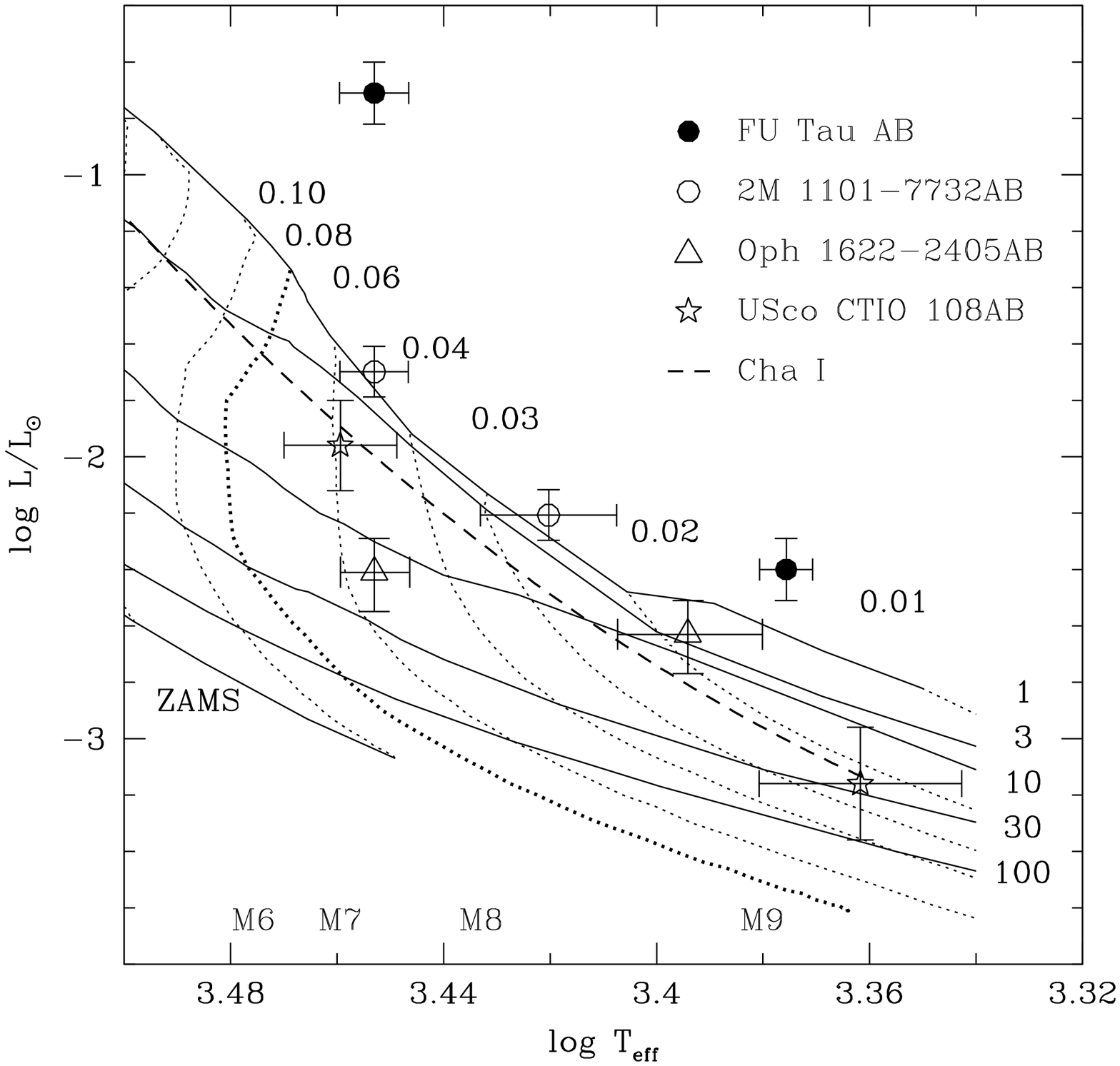}
\caption{
H-R diagram for the components of the binaries FU~Tau ({\it filled circles}, 
Table~\ref{tab:prop}), 2M~J1101$-$7732 \citep[{\it open circles},][]{luh04bin}, 
Oph~1622$-$2405 \citep[{\it triangles},][]{luh07oph}, and
USco~CTIO~108 \citep[{\it stars},][]{bej08}. 
These data are shown with a fit to the empirical cluster sequence for
Chamaeleon~I \citep[{\it dashed line};][]{luh07cha} and the
theoretical evolutionary models of
\citet{bar98} ($0.1<M/M_\odot\leq1$) and \citet{cha00} ($M/M_\odot\leq0.1$),
where the mass tracks ({\it dotted lines}) and isochrones ({\it solid lines})
are labeled in units of $M_\odot$ and Myr, respectively.
}
\label{fig:hr}
\end{figure}

Our temperature and luminosity estimates for FU~Tau~A and B 
are plotted on the H-R diagram in Figure~\ref{fig:hr} with
the evolutionary models from \citet{bar98} and \citet{cha00}.
Although the components of binary systems are expected to be coeval,
FU~Tau~A and B do not appear near the same model isochrone. 
The models imply an age of 1 Myr for the secondary while the primary is 
overluminous by an order of magnitude relative to that isochrone. 
The discrepancy is reduced somewhat if FU~Tau~A and B are compared to a
fit to the cluster sequence for Chamaeleon~I \citep{luh07cha}, which acts 
as an empirical isochrone. As shown in Figure~\ref{fig:hr}, the primary 
and secondary appear 1.3 and 0.6~dex above this fit, respectively,
resulting in a luminosity ratio that is 0.7~dex larger than expected for 
a coeval pair. FU~Tau~A is much brighter than other members of Taurus with 
similar spectral types whereas FU~Tau~B falls within the cluster 
sequence \citep{luh04tau}, indicating that the difference in
isochronal ages for the pair is a reflection of an anomalously high 
luminosity for the primary rather than a low luminosity for the secondary.
One possible explanation is that FU~Tau~A is an unresolved binary. 
If so, it would appear near the upper envelope of the cluster 
sequence for Taurus in the H-R diagram, although it would remain
overluminous relative to FU~Tau~B by $\sim0.4$~dex.

The data and models in Figure~\ref{fig:hr} 
imply a mass of $\sim0.015$~$M_\odot$ for FU~Tau~B. 
The model predictions do not encompass the temperature and luminosity of
FU~Tau~A, and thus do not provide a direct estimate of its mass. 
However, because of the vertical nature of the mass tracks of low-mass
stars and brown dwarfs at young ages ($\tau<10$~Myr), the spectral type
of FU~Tau~A should provide a good indication of its mass regardless of 
luminosity. As shown in Figure~\ref{fig:hr}, the spectral type of FU~Tau~A
combined with the adopted temperature scale and models imply a mass of
$\sim0.05$~$M_\odot$. This estimate is consistent with the fact that FU~Tau~A 
is slightly cooler than 2M~0535$-$0546 (Figure~\ref{fig:spec}), whose 
components have dynamical masses of 0.054 and 0.034~$M_\odot$ \citep{sta06}

To compare the masses and ages of FU~Tau~A and B to those of 
previously known young low-mass binaries, we have included the components
of 2M~J1101$-$7732, Oph~1622$-$2405, and USco~CTIO~108 in Figure~\ref{fig:hr}.
Because the optical spectra of these systems and FU~Tau have been directly 
compared \citep[Figure~\ref{fig:spec};][]{luh07oph,bej08}, 
their relative spectral types should be quite accurate. 
In addition, the spectral types and photometry have been converted to
temperatures and luminosities with the same methods that we have applied 
to FU~Tau. 
As shown in Figure~\ref{fig:hr}, FU~Tau~A probably has a mass that is similar
to those of the other three primaries while FU~Tau~B may be slightly less 
massive than Oph~1622$-$2405~B and USco~CTIO~108~B.
The most striking aspect of this comparison is that the luminosity ratio of 
FU~Tau~A and B ($L_A/L_B\sim80$) is much larger than the ratios in the
other systems ($L_A/L_B=1.6$--16), further illustrating that one of the
components of FU~Tau (probably the primary) has an anomalous luminosity.

\subsection{Spectral Energy Distributions}
\label{sec:sed}

\begin{figure}
\epsscale{1}
\plotone{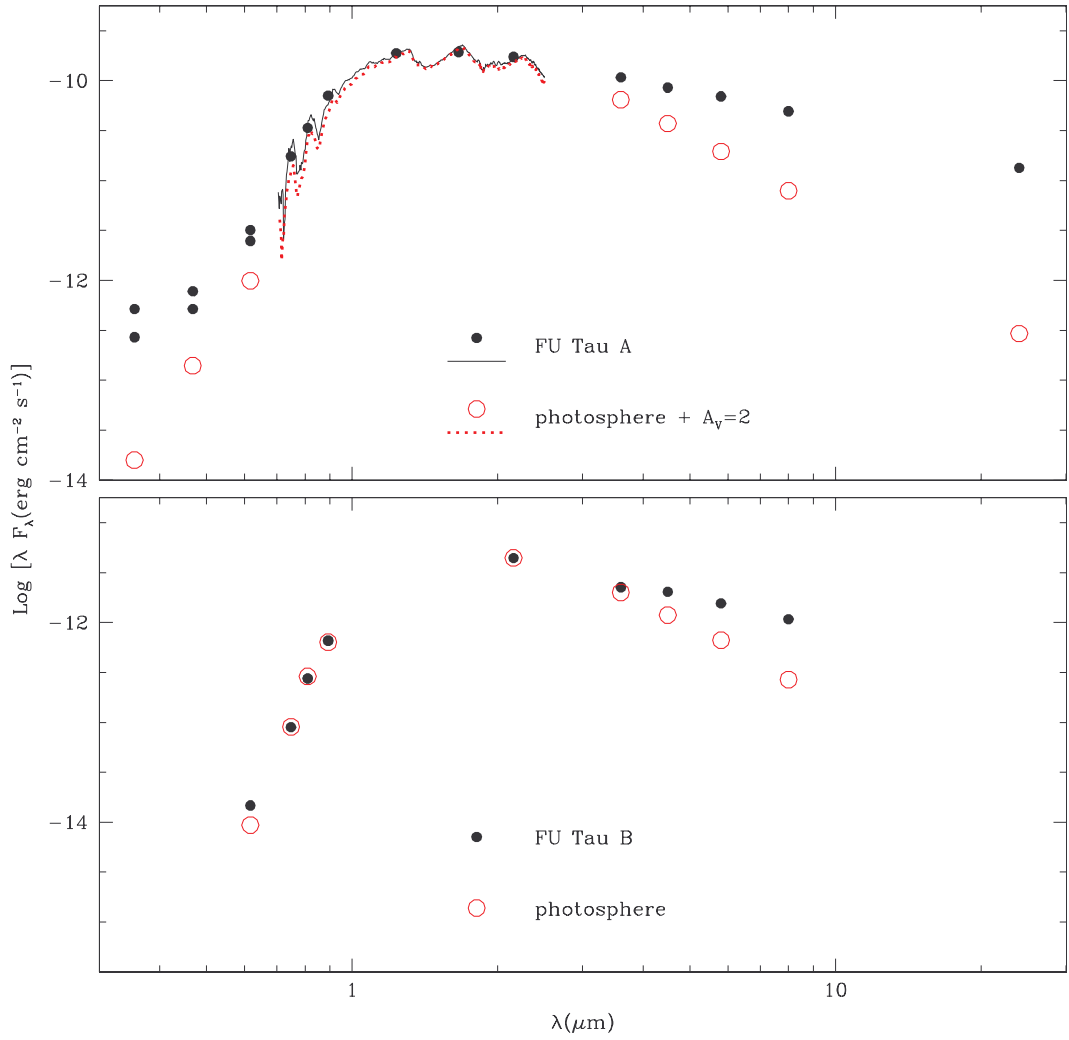}
\caption{
SEDs for FU~Tau~A and B compared to the SEDs of stellar photospheres with
the same spectral types. The two epochs of photometry at $u$, $g$, and $r$ 
(0.35, 0.47, 0.62~\micron) for the primary 
are plotted separately while other data at multiple epochs are averaged. 
The photospheric SEDs have been scaled to the photometry at $J$ and $K$
for the primary and secondary, respectively.
}
\label{fig:sed}
\end{figure}

We selected FU~Tau~A as a candidate member of Taurus because it 
exhibits mid-IR colors that are indicative of a circumstellar disk.
To demonstrate the presence of this mid-IR excess emission from
FU~Tau~A and to determine whether its companion has a disk, we have 
constructed their SEDs using the photometry compiled in Table~\ref{tab:phot}.
SDSS and IRAC have provided photometry at multiple epochs for the 
FU~Tau system; we adopt the average of the available measurements for a given
SDSS band and we use the IRAC data from 2007 since the camera 
observed FU~Tau in only two bands in the other epoch. 
In addition to the photometry, we include in the SED for FU~Tau~A the 
low-resolution IR spectrum obtained with SpeX, which has been 
flux-calibrated with the photometry from 2MASS. 
The resulting SEDs of FU~Tau~A and B are presented in Figure~\ref{fig:sed}.

To determine if the SEDs contain excess emission at short or 
long wavelengths, we compare them to the SEDs of stellar photospheres. 
We construct an estimate of the photospheric SEDs of FU~Tau~A and B
from data for the Taurus members KPNO~5 and KPNO~4, respectively, which 
have similar spectral types as FU~Tau~A and B, have negligible extinction, 
and lack mid-IR excess emission and significant accretion 
\citep{bri02,muz05,har05,luh06tau2}.
Optical measurements in the SDSS and $I$ bands for the KPNO objects are taken 
from \citet{fin04} and \citet{bri02}. Because KPNO 4 has a very low 
signal-to-noise ratio at $r$, we compute the flux in this band by combining 
the $i$ photometry for KPNO~4 with the typical value of $r-i$ for M9 dwarfs 
\citep{kra07}. For KPNO~5, we include a low-resolution IR spectrum collected 
with SpeX \citep{mue07}. 
For the mid-IR portion of these photospheric SEDs, we adopt the average colors 
of diskless stars near the spectral types of FU~Tau~A and B \citep{luh08cha}.  
The template for FU~Tau~A is reddened by $A_V=2$ according to the reddening 
laws from \citet{rl85} and \citet{fla07}.
The photospheric SEDs are then normalized to the $J$- and $K$-band fluxes
of FU~Tau~A and B, respectively. 
Relative to these SEDs, both FU~Tau~A and B exhibit significant excess emission at wavelengths longward of 4~\micron, as shown in Figure~\ref{fig:sed}.
The primary also has excess emission in the bluest optical and UV bands,
which probably explains why its optical spectrum implies less extinction than
the near-IR data (\S~\ref{sec:class}).

To classify the SEDs of FU~Tau~A and B according to the standard scheme
for young stars \citep{lada87,gre94}, we use the IR spectral slope that is
defined as 
$\alpha= d$~log$(\lambda F_\lambda)/d$~log$(\lambda)$ \citep{lw84,adams87}.
The extinction-corrected slopes between 3.6 and 8~\micron\ for each object
are given in Table~\ref{tab:prop}.
By applying the thresholds from \citet{luh08cha} to these slopes,
we classify the SEDs of FU~Tau~A and B as Class~II.

\subsection{Evidence of Membership in Taurus}
\label{sec:mem}

The components of FU~Tau clearly have ages of $\sim1$~Myr based
on their H$\alpha$ emission, mid-IR excess emission, and 
gravity-sensitive spectral features.
Given that they are projected against the center of Barnard 215, 
it is likely that they were born in this cloud, which is a part of the
Taurus star-forming region. The SDSS images of Barnard 215 also show faint
nebulosity centered on FU~Tau, which supports its association with the cloud. 
The proper motion of FU~Tau represents an additional constraint on its
membership.
Indeed, \citet{jh79} identified FU~Tau has a likely member of
Taurus through a measurement of its proper motion. 
We now perform a new analysis of the motion of FU~Tau that makes use
of the astrometric data that are currently available.

In Table \ref{tab:pm}, we list 
proper motion measurements for FU~Tau~A from the USNO-B1.0 catalog
\citep{Monet03} and \citet[][]{Ducourant05}. We also include our
measurement of the proper motion, which is based on positions in the USNO-A2.0 
catalog \citep[][epoch 1950.9]{Monet98}, the Guide Star Catalog V2.3.2
\citep[][epoch 1994.8]{STSci06}, the 2MASS Point Source Catalog \citep[][epoch
1997.9]{skr06}, and the Carlsberg Meridian Catalog Vol. 14 (CDS catalog I/304; 
epoch 2001.8). Our estimate is similar to the value from USNO-B1.0.
We adopt our measurement of the proper motion in the following discussion.

What is the proper motion of a hypothetical member of Taurus at the
position of FU~Tau~A? Because Taurus covers a large solid angle of sky, 
the proper motion for a given velocity vector varies across the 
association due to geometric projection effects.
For this calculation, we adopt a mean Galactic velocity vector 
of (U,V,W)$ = (-16.5, -13.2, -11.0)$~km~s$^{-1}$ \citep{Bertout06} and a 
mean distance of 140~pc for Taurus \citep{wic98,Loinard05,tor07}. 
At the position of FU~Tau~A, we predict that a star
with the mean velocity vector of Taurus would have a proper motion of
$\mu_{\alpha *}=+8.4$~(140~pc/$d$)~mas~yr$^{-1}$ and
$\mu_{\delta}=-23.5$~(140~pc/$d$)~mas~yr$^{-1}$, where $d$ 
is the distance in pc.
The mean velocity vector is constrained to $\pm$1~km~s$^{-1}$
in each component, which translates to an uncertainty of
$\sim$1~mas~yr$^{-1}$ in the predicted proper motion for a given distance. 
Hence, we find that our predicted values of $\mu_{\alpha *}$ and
$\mu_{\delta}$ for an ``ideal'' Taurus member are within 0.3$\sigma$ 
and 1.7$\sigma$ of the observed motion, respectively. The predicted motion 
is also within 1.1~$\sigma$ and 2.4~$\sigma$ of the USNO-B1.0 $\mu_{\alpha *}$ 
and $\mu_{\delta}$ motions, respectively. For comparison, in 
Table \ref{tab:pm} we also include the predicted proper motions for 
hypothetical members of the Hyades and Pleiades open clusters at their 
respective mean distances, which are based on data from
\citet{deBruijne01}, \citet{Robichon99}, and \cite{Soderblom05}. 
Both groups are known to have members in the vicinity of Taurus. However,
the motion of FU~Tau~A is completely inconsistent with 
the kinematic membership of either of these clusters. Based on this analysis,
we conclude that the proper motion of FU~Tau~A is in good agreement with the 
average motion of the stellar population in Taurus. 
According to the astrometric measurements of FU~Tau~A and B in 
Table~\ref{tab:astro}, the pair maintained the same relative positions to 
within $\sim0.1\arcsec$ across a period of three years, indicating that the 
secondary shares the same motion as the primary at a level of 
$\sim$30~mas~yr$^{-1}$.

It is useful to compare the motion of FU~Tau~A to that of FT~Tau, which is
the only other known young star that is projected against the Barnard 215
cloud (Figure~\ref{fig:sdss}).
We have measured the proper motion of FT~Tau using the same catalogs 
that were employed for FU~Tau~A, arriving at values of 
$\mu_{\alpha *}=+6.3\pm3.3$~mas~yr$^{-1}$ and 
$\mu_{\delta}=-15.3\pm3.3$~mas~yr$^{-1}$. 
The motions of the two objects agree to within 1.1$\pm$4.9 and 
2.2$\pm$4.7~mas~yr$^{-1}$ in right ascension and declination, respectively,
which translates into tangential velocities at 140~pc that agree
to within 0.7$\pm$3.3 and 1.5$\pm$3.1~km~s$^{-1}$.
These relative motions indicate that FU~Tau~A has remained within a projected
distance of 1.7~pc ($0.7\arcdeg$) from FT~Tau during its lifetime
($\tau\lesssim1$~Myr, Figure~\ref{fig:hr}). The agreement in their velocities
and the presence of nebulosity centered on each source strongly suggests
that they are associated with the Barnard 215 cloud and were not born elsewhere.

Finally, we have searched for evidence of additional young stars near
Barnard 215. 
Aside from FU~Tau and FT~Tau, the only other candidate young star that has
been previously identified near the cloud is GT~Tau \citep{jh79}.
Through an analysis similar to that performed for FU~Tau and FT~Tau, 
we find that the proper motion of GT~Tau is inconsistent with membership
in Taurus. Previous surveys for young stars in the vicinity of Barnard 215 have
been conducted at optical wavelengths, and thus could have missed
stars that are embedded within the cloud. 
The mid-IR images from {\it Spitzer} that were used to uncover FU~Tau
easily penetrate the extinction of Barnard 215, and they do not reveal 
any additional disk-bearing stars. We cannot rule out the presence
of diskless young stars within the cloud given that they would be
indistinguishable from field stars in {\it Spitzer} data. 
However, it is unlikely that an embedded population within the cloud would 
consist of only diskless stars. 

\subsection{Evidence of Binarity}
\label{sec:bin}

We have presented strong evidence indicating that FU~Tau~A and B are 
members of the Taurus star-forming region.
We have also shown that they have similar proper motions, but the 
accuracy of these measurements is insufficient to distinguish
between a binary system and a pair of unrelated members of Taurus that 
are seen in projection near each other. 
Therefore, as in most studies of visual pairs in star-forming regions, 
we rely on a statistical analysis to assess whether FU~Tau~A and B
are likely to comprise a binary system. 
Because of the low stellar density in Taurus, it is unlikely that a
given member of the region will appear close to another member on the sky,
unless they are in a binary system.
This probability is particularly low for FU~Tau, which resides in 
one of the most isolated dark clouds in Taurus.
As shown in Figure~\ref{fig:map}, only one young star has
been found within a radius of $0.5\arcdeg$ surrounding FU~Tau.
We can roughly quantify the probability that FU~Tau is a pair of unrelated
low-mass Taurus members.
Among the 51 known members of Taurus that have spectral types later than M6,
the median distance to the nearest $>$M6 neighbor is $22.3\arcmin$.
For our probability calculation, we adopt a population of 51 objects
that are randomly distributed across an area of 32.5~deg$^{2}$, which
exhibits a median nearest neighbor distance that is similar to the observed
value. The probability of finding a pair of objects in this population
with a projected separation of $a\leq6\arcsec$ is $3\times10^{-4}$.
Therefore, it is likely that FU~Tau~A and B comprise a binary system.

\section{Discussion}
\label{sec:disc}

With component masses of $\sim$0.05 and $\sim$0.015~$M_\odot$ and a projected 
separation of 800~AU, FU~Tau joins the growing sample of wide low-mass binaries
that have been discovered in recent years (\S~\ref{sec:intro}). 
These systems have provided valuable constraints on theories of the
formation of brown dwarfs \citep{luh04bin}. 
Unlike the previously known binaries -- or free-floating brown dwarfs, for that
matter -- FU~Tau~A and B are the first spectroscopically-confirmed
brown dwarfs discovered forming in isolation\footnote{
FU~Tau may comprise a more evolved version of L1014-IRS, which is an
isolated low-mass protostar \citep{you04,bou05,hua06}. Its current and 
ultimate masses are uncertain because it is extremely young and highly
embedded.}. The unique formation environment of FU~Tau represents a new 
test of theoretical models for the origin of brown dwarfs.

As reviewed in \S~\ref{sec:intro}, dynamical models 
typically create brown dwarfs through ejection from protostellar clusters 
\citep{rc01,bat02} or fragmentation of disks around stars and subsequent
stripping of the substellar companion through encounters with other stars
\citep{goo07,sta07}.
Because both mechanisms require the presence of stellar clusters, 
they cannot account for the existence of brown dwarfs like FU~Tau~A and B
that are forming in isolation. 
Although the disk fragmentation models produce brown dwarfs as companions
to stars, and those stars can be isolated, this scenario is not 
plausible for FU~Tau because the mass and separation of the secondary
are much larger than the (limited) measurements of masses and radii for brown 
dwarf disks \citep[$r\sim30$~AU, 
$M\sim1$~$M_{\rm Jup}$;][]{kle03,sch06,luh07edgeon}.

The discovery of FU~Tau demonstrates that brown dwarfs can arise in isolation
without the involvement of dynamical interactions among stars.
It is likely that the same process that made it possible for substellar
objects to form in Barnard 215 also occurs in clusters as well.
Thus, it is unnecessary to invoke additional mechanisms like ejection
for creating brown dwarfs unless observations indicate their presence.
Dynamical interactions in young clusters probably do affect the formation of 
brown dwarfs, just as they affect the formation of stars, but there is 
no evidence to date that they represent a second dominant mechanism that 
enables the birth of brown dwarfs \citep{luh07ppv}.

\acknowledgements

We thank Lee Hartmann for helpful comments on the manuscript.
K.~L. was supported by grant AST-0544588 from the National Science 
Foundation (NSF) and E.~M. was supported by a Clay Postdoctoral Fellowship from 
the Smithsonian Astrophysical Observatory. This publication makes use of data 
products from 2MASS, which is a joint project of the University of 
Massachusetts and the Infrared Processing and Analysis Center/California 
Institute of Technology, funded by NASA and the NSF. The Guide Star 
Catalogue-II is a joint project of the Space Telescope Science Institute and
the Osservatorio Astronomico di Torino. Space Telescope Science Institute is 
operated by the Association of Universities for Research in Astronomy, for NASA
under contract NAS5-26555. The participation of the Osservatorio Astronomico di
Torino is supported by the Italian Council for Research in
Astronomy. Additional support is provided by European Southern
Observatory, Space Telescope European Coordinating Facility, the International 
GEMINI project and the European Space Agency Astrophysics Division.
This research used the facilities of the Canadian Astronomy Data Centre 
operated by the National Research Council of Canada with the support of the
Canadian Space Agency.
Funding for SDSS has been provided by the Alfred P. Sloan Foundation, 
the Participating Institutions, the NSF,
the U.S. Department of Energy, NASA, the Japanese Monbukagakusho, the 
Max Planck Society, and the Higher Education Funding Council for England. 
The SDSS is managed by the Astrophysical Research Consortium for the 
Participating Institutions. The Participating Institutions are the American 
Museum of Natural History, Astrophysical Institute Potsdam, University of 
Basel, University of Cambridge, Case Western Reserve University, The University 
of Chicago, Drexel University, Fermilab, the Institute for Advanced Study, the 
Japan Participation Group, The Johns Hopkins University, the Joint Institute 
for Nuclear Astrophysics, the Kavli Institute for Particle Astrophysics and 
Cosmology, the Korean Scientist Group, the Chinese Academy of Sciences, 
Los Alamos National Laboratory, the Max-Planck-Institute for Astronomy, 
the Max-Planck-Institute for Astrophysics, New Mexico State University, 
Ohio State University, University of Pittsburgh, University of Portsmouth, 
Princeton University, the United States Naval Observatory, and the University 
of Washington.

\clearpage

\begin{deluxetable}{llll}
\tabletypesize{\scriptsize}
\tablewidth{0pt}
\tablecaption{Photometry for FU Tau A and B\label{tab:phot}}
\tablehead{
\colhead{Band} &
\colhead{A} &
\colhead{B} &
\colhead{Date}}
\startdata
$u$ & 20.14$\pm$0.06 & \nodata & 2002 Dec 6 \\
$u$ & 19.43$\pm$0.05 & \nodata & 2002 Dec 29 \\
$g$ & 19.13$\pm$0.05 & \nodata & 2002 Dec 6 \\
$g$ & 18.69$\pm$0.05 & \nodata & 2002 Dec 29 \\
$r$ & 17.13$\pm$0.05 & 22.6$\pm$0.3 & 2002 Dec 6 \\
$r$ & 16.86$\pm$0.05 & 22.8$\pm$0.3 & 2002 Dec 29 \\
$i$ & 14.86$\pm$0.05 & 20.45$\pm$0.06 & 2002 Dec 6 \\
$i$ & 14.75$\pm$0.05 & 20.62$\pm$0.06 & 2002 Dec 29 \\
$z$ & 13.12$\pm$0.05 & 18.23$\pm$0.05 & 2002 Dec 6 \\
$z$ & 13.05$\pm$0.05 & 18.11$\pm$0.05 & 2002 Dec 29 \\
$I$ & 13.58$\pm$0.04 & 18.80$\pm$0.05 & 2002 Dec 29 \\
$J$ & 10.78$\pm$0.02 & \nodata & 1997 Nov 29 \\
$H$ & 9.95$\pm$0.03 & \nodata & 1997 Nov 29 \\
$K_s$ & 9.32$\pm$0.02 & \nodata & 1997 Nov 29 \\
$K$ & saturated & 13.33$\pm$0.02 & 2005 Dec 31 \\
$[3.6]$ & out & out & 2005 Feb 23 \\
$[3.6]$ & 8.34$\pm$0.02 & 12.54$\pm$0.1 & 2007 Mar 3 \\
$[4.5]$ & 7.68$\pm$0.02 & 11.90$\pm$0.1 & 2005 Feb 23 \\
$[4.5]$ & 7.87$\pm$0.02 & 11.93$\pm$0.1 & 2007 Mar 3 \\
$[5.8]$ & out & out & 2005 Feb 23 \\
$[5.8]$ & 7.33$\pm$0.03 & 11.46$\pm$0.1 & 2007 Mar 3 \\
$[8.0]$ & 6.45$\pm$0.03 & 10.77$\pm$0.1 & 2005 Feb 23 \\
$[8.0]$ & 6.72$\pm$0.03 & 10.87$\pm$0.1 & 2007 Mar 3 \\
$[24]$ & 4.56$\pm$0.04 & \nodata & 2007 Feb 2 \\
\enddata
\tablecomments{Data are from SDSS ($ugriz$), CFHT ($I$), 2MASS ($JHK_s$),
UKIDSS ($K$), and {\it Spitzer} (3.6--24~\micron).}
\end{deluxetable}

\begin{deluxetable}{llllll}
\tabletypesize{\scriptsize}
\tablewidth{0pt}
\tablecaption{Astrometry for FU Tau A and B\label{tab:astro}}
\tablehead{
\colhead{} &
\colhead{Separation} &
\colhead{PA} &
\colhead{} \\
\colhead{Source} &
\colhead{(arcsec)} &
\colhead{(deg)} &
\colhead{Date}}
\startdata
CFHT & 5.61$\pm$0.1 & 123.0$\pm$1 & 2002 Dec 29 \\
SDSS & 5.67$\pm$0.1 & 123.4$\pm$1 & 2002 Dec 6, 29 \\
UKIDSS & 5.72$\pm$0.1 & 123.2$\pm$1 & 2005 Dec 31 \\
\enddata
\end{deluxetable}

\begin{deluxetable}{llllllll}
\tabletypesize{\scriptsize}
\tablewidth{0pt}
\tablecaption{Properties of FU Tau A and B\label{tab:prop}}
\tablehead{
\colhead{} &
\colhead{} &
\colhead{$T_{\rm eff}$\tablenotemark{a}} &
\colhead{} &
\colhead{$L_{\rm bol}$} &
\colhead{Membership} &
\colhead{$W_{\lambda}$(H$\alpha$)} &
\colhead{} \\ 
\colhead{FU Tau} &
\colhead{Spectral Type} &
\colhead{(K)} &
\colhead{$A_V$} &
\colhead{($L_\odot$)} &
\colhead{Evidence\tablenotemark{b}} &
\colhead{(\AA)} &
\colhead{$\alpha$(3.6-8~\micron)} 
} 
\startdata
A & M7.25$\pm$0.25 & 2838 & 2 & 0.19 & NaK,H$_2$O,ex,e,$\mu$ & 93$\pm$7 & $-$1.02 \\
B & M9.25$\pm$0.25 & 2375 & $<1$ & 0.0039 & NaK,ex,e & $\sim$70 & $-$0.92 \\
\enddata
\tablenotetext{a}{Temperature scale from \citet{luh03ic}.}
\tablenotetext{b}{Membership in Taurus is indicated by strong emission 
lines (``e"), Na~I and K~I strengths intermediate between those of dwarfs 
and giants (``NaK"), the shape of the gravity-sensitive steam bands (``H$_2$O"),
IR excess emission (``ex"), or a proper motion (``$\mu$")
that is similar to that of the known members of the star-forming region.}
\end{deluxetable}

\begin{deluxetable}{llllllll}
\tabletypesize{\scriptsize}
\tablewidth{0pt}
\tablecaption{Observed and Predicted Proper Motions of FU Tau A\label{tab:pm}}
\tablehead{
\colhead{} & 
\multicolumn{3}{c}{Observed} &
\colhead{} &
\multicolumn{3}{c}{Predicted} \\
\cline{2-4} \cline{6-8}
\colhead{} & 
\colhead{USNO-B1.0\tablenotemark{a}} &
\colhead{Duc05\tablenotemark{b}} &
\colhead{this work} &
\colhead{} &
\colhead{Taurus} &
\colhead{Hyades} &
\colhead{Pleiades}} 
\startdata
$\mu_{\alpha *}$  & +6\,$\pm$\,2 & +14\,$\pm$\,7 & +7.2\,$\pm$\,3.6 & & +8.4 & +109.3 & +14.5 \\
$\mu_{\delta}$  & $-$18\,$\pm$\,2 & $-$26\,$\pm$\,7 & $-$17.5\,$\pm$\,3.4 & & $-$23.5 & $-$52.9 & $-$47.2 \\
\enddata
\tablecomments{Units are mas\,yr$^{-1}$.}
\tablenotetext{a}{\citet{Monet03}.}
\tablenotetext{b}{\citet{Ducourant05}.}
\end{deluxetable}

\clearpage

\end{document}